\documentclass[trackchanges]{aastex701}

\usepackage{xcolor}
\usepackage{amsmath}
\usepackage{amssymb}
\usepackage{wrapfig}

\begin{document}

\title{Plasma Heating and Energization in Hot-Onset Flare Precursor Events}


\author[orcid=0009-0007-0123-9821,sname='Da']{H. Da}
\affiliation{IREAP, University of Maryland}
\email[show]{dht@umd.edu}  

\author[orcid=0000-0002-9150-1841,sname='Drake']{J. F. Drake}
\affiliation{Dept. of Physics, Institute for Physical Science and Technology, and the Joint Space Science Institute, University of Maryland}
\email[show]{drake@umd.edu}

\author[orcid=0000-0002-5435-3544,sname='Swisdak']{M. Swisdak}
\affiliation{IREAP, University of Maryland}
\email[show]{swisdak@umd.edu} 








\begin{abstract}

Recent observations of solar flares have revealed a distinct ``hot onset'' phase preceding the impulsive phase, characterized by elevated plasma temperature ($>10$ MK), a steady increase in emission measure, and the absence of detectable nonthermal emission. The standard flare model attributes the formation of hot flare plasma to chromospheric evaporation driven by energy deposition from nonthermal electrons, and therefore does not naturally explain why thermal heating precedes nonthermal emission. Motivated by recent observations of guide-field evolution in two-ribbon flares and by large-scale MHD simulations, we investigate hot-onset energization using simulations of magnetic reconnection with the kinetic macroscale model {\it kglobal} initialized with a spatially non-uniform guide field. Because the efficiency of electron non-thermal energy production from Fermi acceleration decreases with guide-field strength, the evolving guide field produces two distinct energization regimes. During the initial high-guide-field phase, the bulk electron temperature rises to approximately twice its initial value, while non-thermal particle production remains weak. Upon entering the low-guide-field regime, energetic particle production is significantly enhanced, with the population of high-energy electrons growing rapidly. These findings provide a self-consistent physical explanation for the hot onset, demonstrating that the macroscopic evolution of the guide field naturally bridges the hot-onset and impulsive phases of solar flares without requiring a separate trigger mechanism.

\end{abstract}



\section{Introduction} 

Solar flares, intense eruptions of electromagnetic radiation that 
last from minutes to hours, are  frequently divided into pre-flare, impulsive, and decay phases. During the impulsive phase, magnetic reconnection transfers energy from the magnetic field into the thermal and non-thermal energy of charged particles. In the standard CSHKP flare model \citep{Carmichael64,sturrock66,Hirayama74,Kopp76}, filament channels above the solar chromosphere become unstable and the magnetic field lines stretch, forming a thin current sheet in the corona separating magnetic fields with anti-parallel components. Then, magnetic reconnection triggered in the current layer heats in-flowing plasma and accelerates particles to high energy. In recent simulations, the reconnection process drives the production of both hot thermal and non-thermal distributions of electrons and protons \citep{arnold21a}. Non-thermal electrons follow field lines down to the chromosphere where they collide with high-density plasma at the footpoints and emit hard X-rays. The hot plasma from the acceleration region and chromospheric evaporation combine to produce soft X-rays. The existence of thermal and non-thermal energization during reconnection is supported by solar observational data \citep{warmuth16b} as well as {\it in situ} measurements in the Earth's magnetotail \citep{ergun20b,rajhans25}. 

Pre-flare solar activity has been observed for more than forty years \citep{benz83}, but a recent study of pre-flare events revealed that there is an extended "hot-onset" interval where the intensity of soft X-rays  gradually increases while plasma temperatures elevate (to $>10$ MK) and remain relatively stationary until the impulsive phase begins \citep{hudson21}. Observations from GOES, RHESSI, and Solar Orbiter/STIX confirm these findings while also demonstrating that the emission measure increases steadily during the hot-onset period, indicating a continuous process of energy release \citep{Battaglia23}. Since non-thermal hard X-ray emission is detected only later in the flare, after an extended period of soft X-ray emission, non-thermal electrons cannot be the dominant driver of hot thermal plasma and associated soft X-rays during the hot-onset phase. The observations suggest that magnetic reconnection during the hot-onset phase drives hot thermal electrons but very few non-thermal electrons, which is not consistent with the standard CSHKP model. Thus, studying the heating mechanism during hot onset will provide valuable insight into the energy transfer and magnetic field variation before the impulsive phase, thereby refining our understanding of flare initiation and plasma energization.

Simulations and modeling suggest that Fermi reflection during flux rope contraction and merging is a strong driver of thermal and non-thermal heating of particles during magnetic reconnection \citep{drake06a,oka10a,drake13a,dahlin14a,li19a,zhang21a}. The efficiency of the non-thermal energization is high when the guide field (the component of the magnetic field perpendicular to the plane of reconnection) is small compared with the asymptotic reconnecting field. In a strong-guide-field environment, non-thermal energy production through Fermi reflection is suppressed, since the curvature at the ends of the magnetic islands is reduced \citep{dahlin14a,dahlin16a}. Reconnection simulations produce harder electron power laws (that is, more non-thermal particles) when the guide field is relatively weak, while electron thermal heating during reconnection is insensitive to the guide field \citep{arnold21a}.
As a result, the strength of the guide field determines whether thermal heating or non-thermal acceleration dominates energization. This suggests a possible explanation of the hot-onset phenomenon: Flares begin during strong-guide-field reconnection that produces only hot thermal plasma and then evolves into a weak-guide-field regime where non-thermal acceleration occurs. 


A challenge to this picture is that direct measurements of the magnetic field structure of the corona are not yet available.  However, the evolution of the guide field has been studied in two-ribbon solar flares and their associated  polarity inversion lines (PILs). Reconnection in two-ribbon flares occurs between radial magnetic fields that have opposite signs on the two ribbons. At the same time, there is a guide-field component oriented along the ribbons. Three-dimensional MHD simulations have revealed the dynamics of reconnection and guide field evolution in two-ribbon flares \citep{dahlin22a}. During the simulation, the guide field is initially large. As the flare progresses, the magnetic field expands to form a vertical current sheet with reconnection occurring between the radial magnetic field components.  The guide field is carried radially outward in a large-scale coronal mass ejection leading to nearly anti-parallel (low-guide-field) reconnection at late time.  Observations from AIA and RHESSI of an M6.9 two-ribbon solar flare showed an example of the evolution of the magnetic shear  \citep{qiu23a} that supports the strong-to-weak guide field transition.  Emission from the heated plasma revealed the evolution of the tilt angle of the post-reconnection flare loops (PRFLs) and, therefore, the strength of the guide field. 
PRFLs tilted along (perpendicular to) the ribbons indicated a strong (weak) guide field. In this event, the guide field weakened prior to the onset of the impulsive phase of the flare. These observations support the picture that the guide field controls the time evolution of particle heating and energization in flares. The reduction of the guide field shifts the system from being particle-heating-dominated to non-thermal-energization dominated, where non-thermal, high-energy electrons are efficiently produced. This behavior parallels what is observed in hot-onset events. Therefore, we suggest that hot-onset events are a consequence of the evolution of the guide field during solar flare reconnection. 

To investigate whether reconnection with a variable guide field can play a significant role during the hot-onset stage of solar flares, we perform numerical simulations using the {\it kglobal} model \citep{drake19a,arnold19a,arnold21a,yin24a,yin24b}.  The model is designed to explore the self-consistent production of energetic particles during magnetic reconnection in macroscale systems. It combines magnetohydrodynamics (MHD) with a macroparticle description, in which energetic particles are treated using guiding-center dynamics and interact self-consistently with the MHD fluid through their pressure anisotropy. Unlike particle-in-cell approaches, {\it kglobal} orders out kinetic scales such as the Debye length or the electron inertial length, enabling simulations of reconnection and particle energization in large-scale systems relevant to solar flares. 

\section{Simulation Model Setup}\label{sec:simulation}

\begin{figure*}[ht!]
\plotone{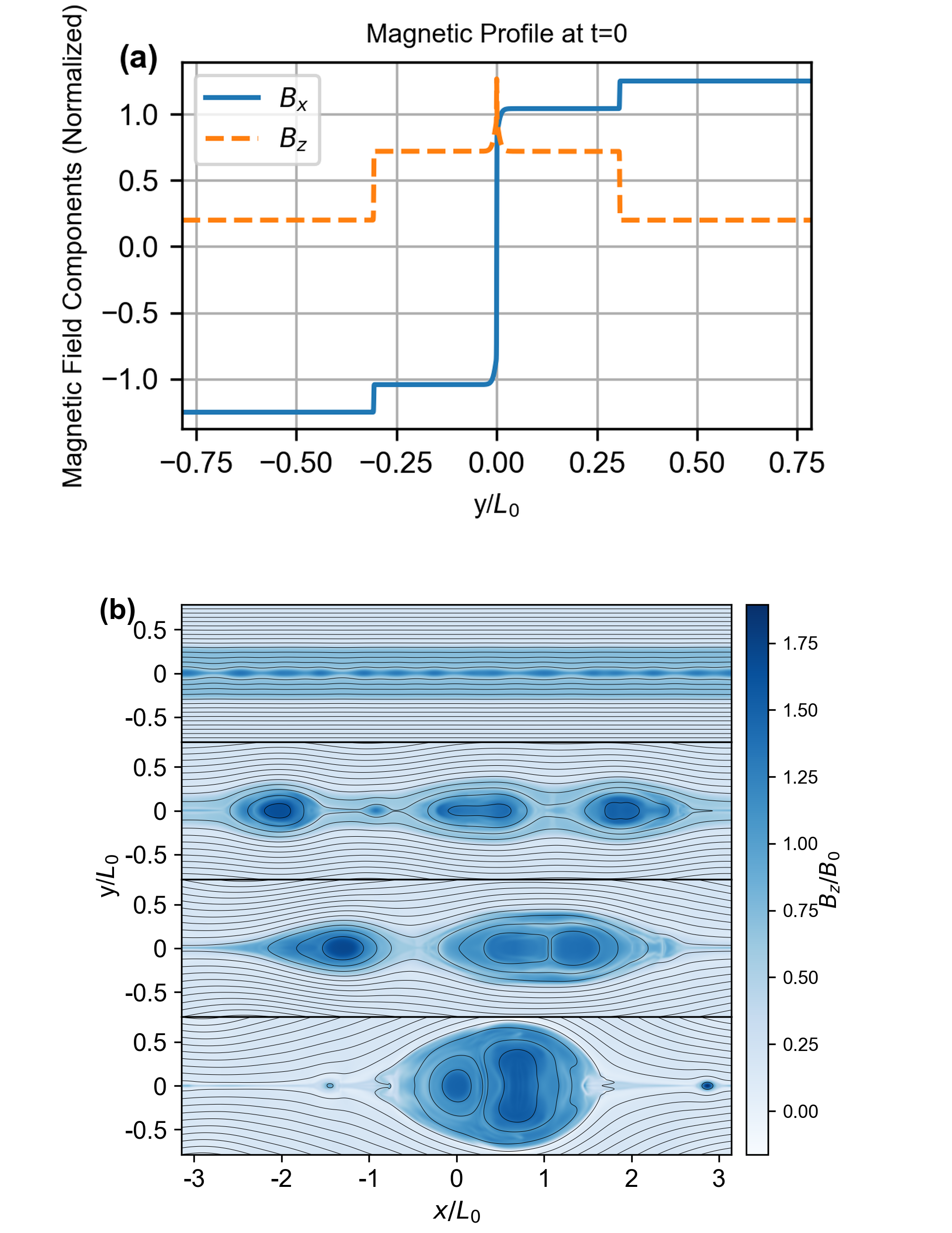}
\caption{(a):Initial profiles of the reconnecting component ($B_x$) and the guide field component ($B_z$) of the magnetic field as a function of $y$. (b):The evolution of the guide field $B_z$ and the formation of islands overplotted with magnetic field lines at $t/\tau_A = 4.5,9.5,13.5,17$. 
\label{fig:profile}}
\end{figure*}

The simulations include four distinct plasma species: fluid protons and electrons, and particle protons and electrons represented by macro-particles that move through the simulation grid. The particle populations are treated in the guiding-center approximation, thereby eliminating the need to resolve their respective Larmor radii. The computational domain measures $L_x \times L_y = 2 \pi L_0 \times \pi L_0$ with $N_x \times N_y=4096 \times 2048$ grid points and employs fully periodic boundary conditions. The initial state includes two current sheets and magnetic field components in both the in-plane ($x$) and out-of-plane ($z$) directions, as shown in Figure \ref{fig:profile} (a). The initial plasma density, temperature, and total magnetic pressure are constant for this equilibrium.  This magnetic field profile includes a thin current sheet at $y=0$ that triggers reconnection. Inflowing plasma carries the upstream guide field into the center of the current sheet so that the spatial structure of the guide field, from the center of the sheet outward, leads to reconnection that mimics the expected evolution in time of the guide field during hot onset. Early in the simulation, when the guide field is about 0.75 of the reconnecting field, thermal heating should dominate. As the simulation develops, the guide field will drop to 0.2 at the current sheet and non-thermal energization should onset.

\begin{figure*}[ht!]
\plotone{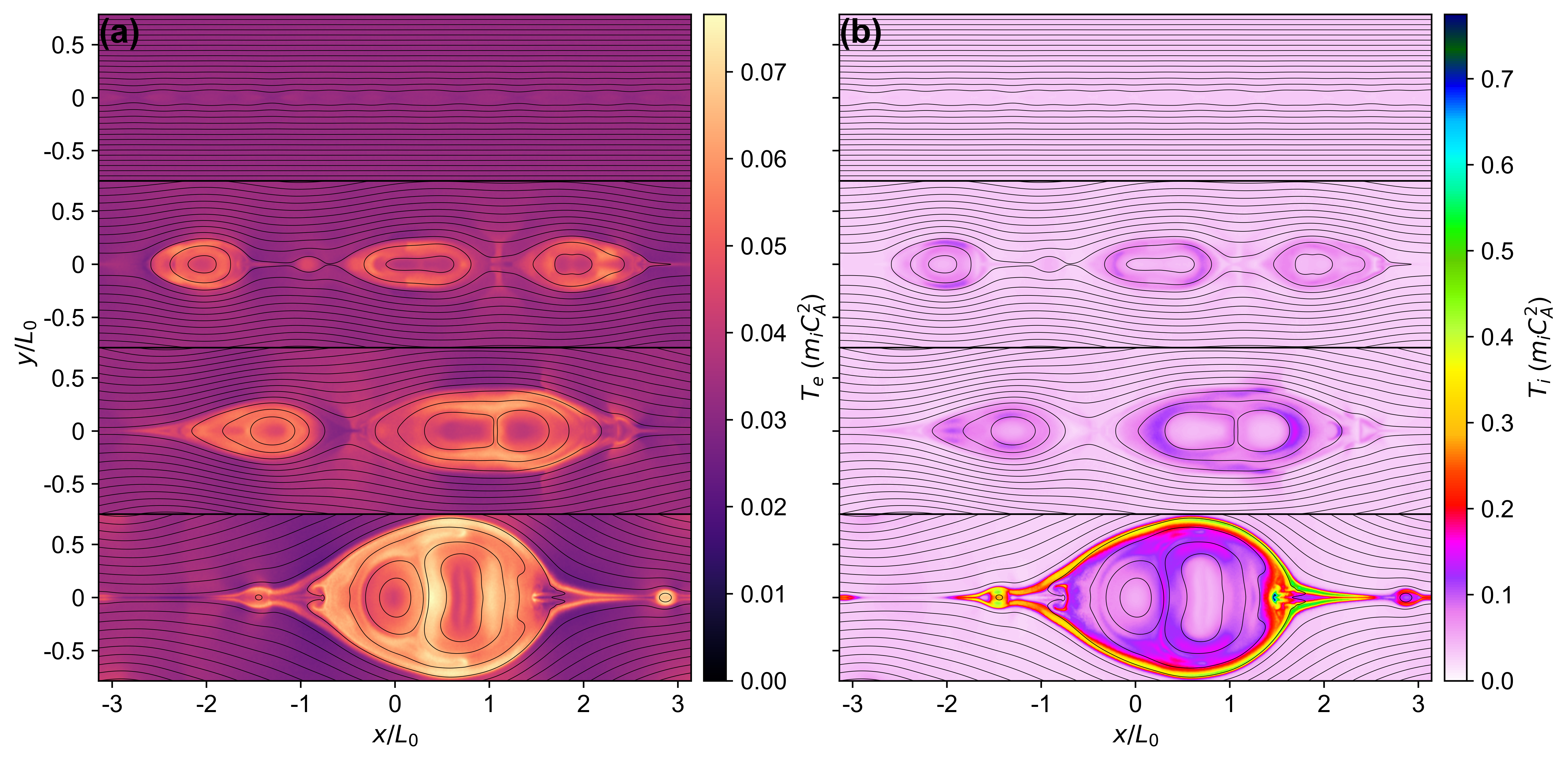}
\caption{In (a) : The evolution of the particle electron temperature $T_e$. In (b): The evolution of the proton particle  temperature $T_i$ at $t/\tau_A = 4.5,9.5,13.5,17$. 
\label{fig:te_ti_2cols_tight}}
\end{figure*}

\begin{figure*}[ht!]
\plotone{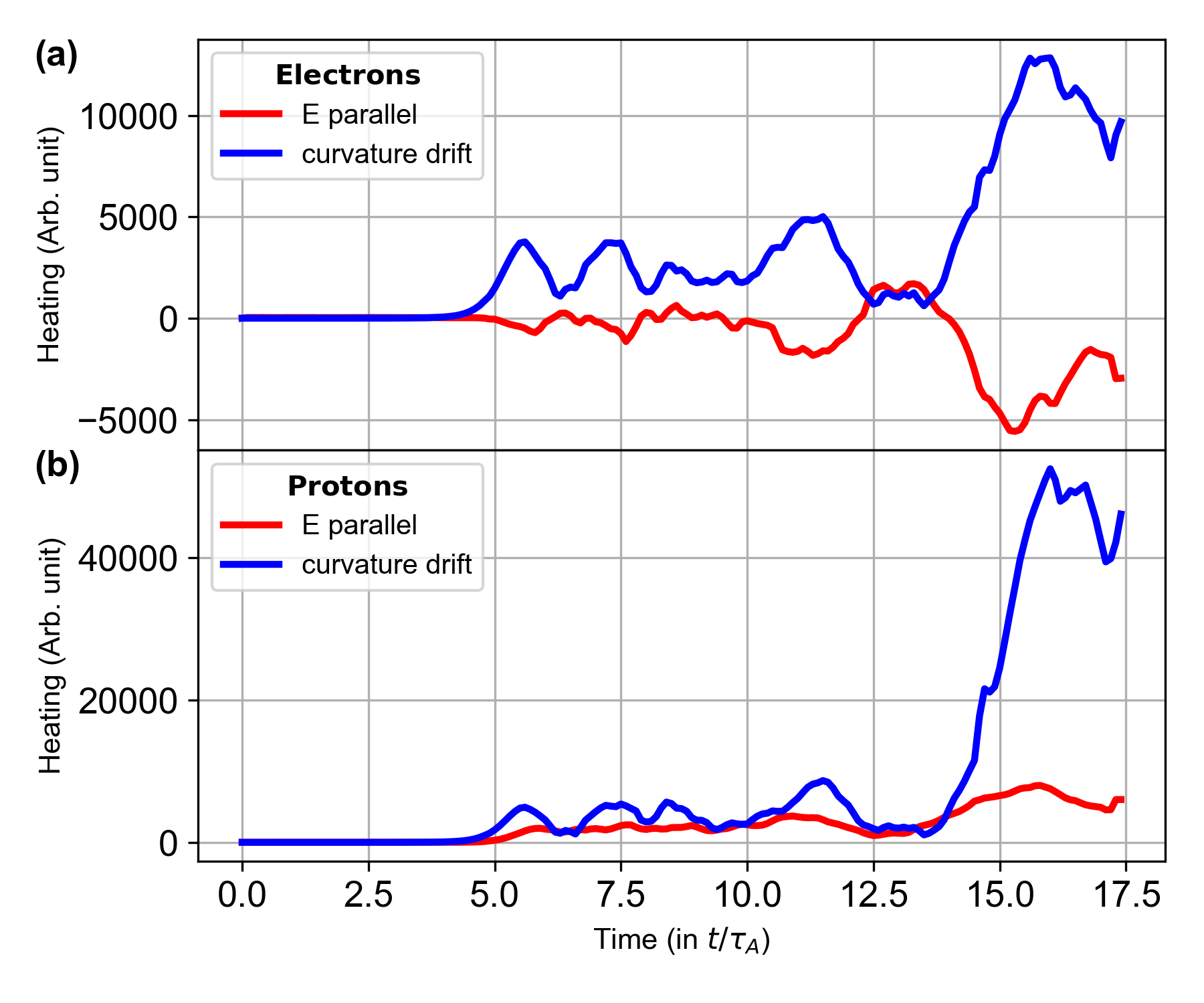}
\caption{ The heating  rate by different sources integrated over the entire domain for both electrons (a) and protons (b). From  Equation~(\ref{eq:electron_energy_equation}); the heating rate from the parallel electric field (red), the curvature drift (blue).
\label{fig:HeatingPower}}
\end{figure*}

\begin{figure*}[ht!]
\plotone{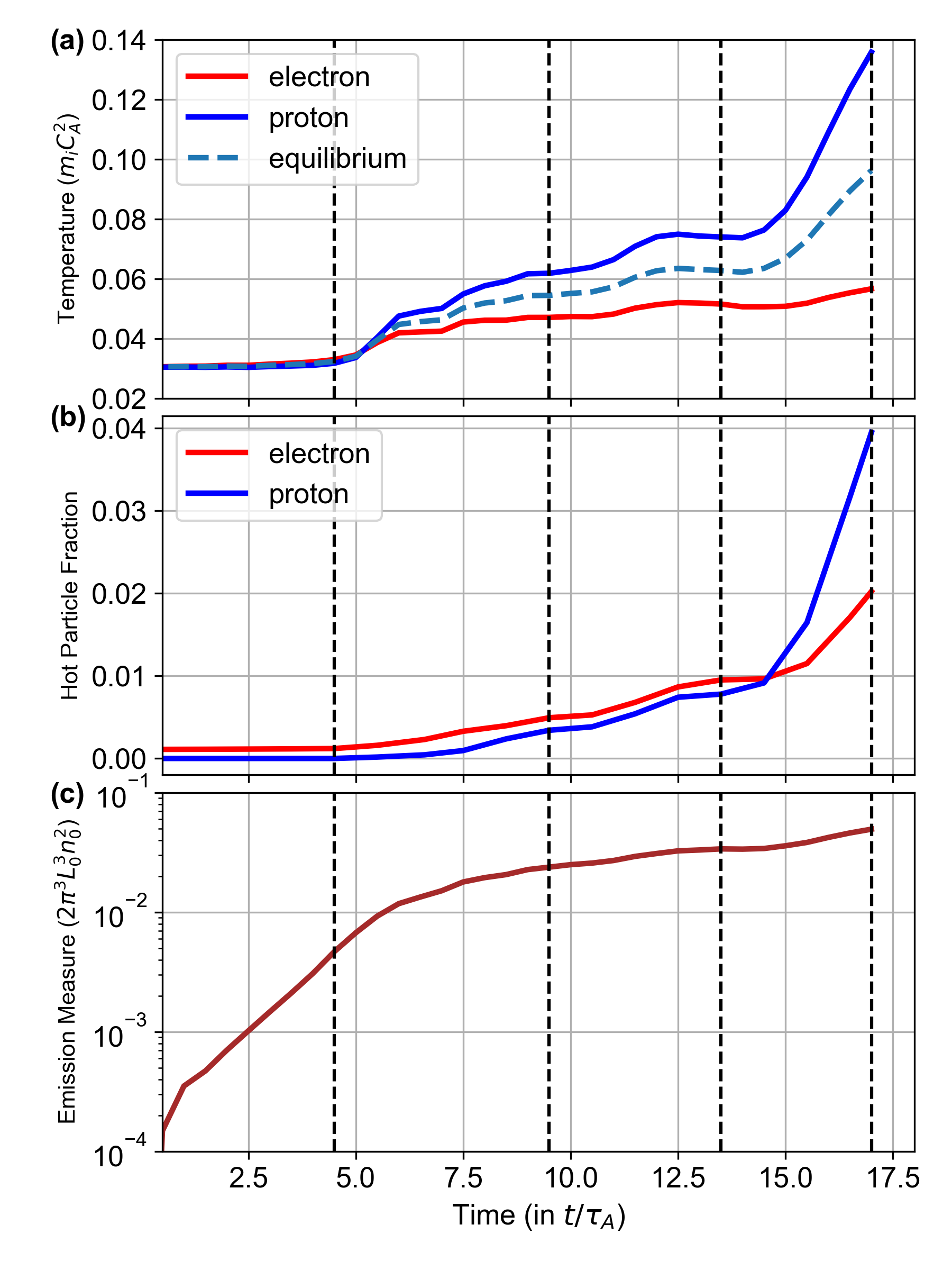}
\caption{(a): The average temperature of electrons and protons within the outermost magnetic separatrix versus time. (b): The number fraction of electrons and protons that have energy exceeding 3 times of the late-time peak thermal energy of the bulk plasma (the last data point in (a) for each of two species). (c) The emission measure, which is integrated over the hot electrons within the outmost magnetic separatrix. The dashed lines are drawn at the four times shown in Figure 2. 
\label{fig:emission_stack}}
\end{figure*}

Since the {\it kglobal} model, like MHD,  does not have an intrinsic spatial scale, we can normalize all spatial variables to an arbitrary macroscale $L_0$; here the initial separation of the current layers is $\pi L_0/2$.  The upstream reconnection magnetic field $B_0$ and the total proton density (the combined number density of particle and fluid protons) $n_{i0}$ define the Alfv\'en speed $C_{A0} = B_0 / \sqrt{4 \pi m_i n_{i0}}$, which is the velocity normalization. The speed of light in the simulation is $\approx30 C_{A0}$.   Time scales are normalized to $\tau_A = L_0 / C_{A0}$. Both temperatures and the particle energies are normalized to $m_i C_{A0}^2$.  For typical solar flare parameters, the energy unit conversion is:
\begin{equation}
m_i C_{A0}^2 \approx 18\text{ keV}\, \left(\frac{B}{60 \text{ G}}\right)^2 \, \left(\frac{10^{10}\text{cm}^{-3}}{n_i}\right).
\label{eq:unit}
\end{equation}
The electron-to-proton mass ratio is taken to be 0.04. The rates of electron heating and energization are insensitive to this value \citep{yin25a}. For this hybrid system, the initial particle proton density is $0.25n_{i0}$ and the initial fluid proton density is $0.75n_{i0}$; the same partition is applied to electrons. The simulations include both species with equal upstream temperatures: $T_e=T_i=0.03125 \,  m_i C_{A0}^2$ and Maxwellian distributions.  The magnetic field evolution equation includes a hyper-resistivity $\nu$ to trigger reconnection, which minimizes dissipation at large scales compared with a standard resisitivity. The effective  Lundquist number is then $S_\nu = C_A L_0^3 / \nu=3.76*10^7$, which is controlled by the size of the computational domain.

\section{ Result} \label{sec:results}

Figure \ref{fig:profile}(b) shows the temporal evolution of the guide field in the plane of the simulation. Several islands form soon after reconnection begins and then merge until only one large island remains. In the early stage (first panel, $t/\tau_A = 4.5$), newly reconnected field lines are associated with a strong guide field. As time progresses, the imposed high guide-field ($B_g = 0.75$) is advected into the plasmoids. At late times, the low–guide-field region reaches the reconnecting layer, as shown in the final panel. This guide-field evolution is a consequence of the prescribed guide field profile shown in Figure~\ref{fig:profile}(a) and is designed to mimic the transition from flare onset to the impulsive phase seen in 3D MHD simulations \citep{dahlin22a}. 
We calculate the particle temperatures from the parallel and perpendicular pressures, which are computed from the particle moments,

\begin{equation}
T_{s}=\frac{2p_{\perp,s}+p_{\parallel,s}}{3\,n_s}, \qquad s\in\{e,i\},
\label{eq:Tiso}
\end{equation}
where $p_{\perp,s}$, $p_{\parallel,s}$ and $n_s$ are the perpendicular and parallel particle pressures and particle density for each species. 
Figure \ref{fig:te_ti_2cols_tight} shows the electron (a) and proton (b) temperatures at several representative times. At $t/\tau_A = 4.5$, reconnection has not yet commenced, and both species maintain their initial uniform temperature of $0.03125 \, m_i C_A^2$ across the entire domain. By $t/\tau_A = 9.5$ and $t/\tau_A = 13.5$, significant heating has taken place over a broad region inside the magnetic islands. Although Fermi reflection is suppressed by the strong guide field, it remains strong enough to produce sustained, spatially extended thermal heating. This behavior is observed for both electrons and protons.

At the later time, $t/\tau_A = 17$, the reduced guide-field strength in the acceleration region significantly enhances Fermi reflection. 
Hot particles concentrated at the edges of the island arise from Fermi reflection with a weak guide field. The heating in the island core is likely driven by island merging. Similar to electrons, proton energization is highly sensitive to the local guide-field strength. In regions of weak guide field, protons undergo strong Fermi acceleration. Counter-streaming ion beams accelerated from opposite ends of the island lead to a pronounced enhancement of the proton temperature. The protons in the outermost shell experience multiple small-island mergers near the x-line where the guide field in the acceleration region has dropped. They gain substantially more energy than protons that undergo multiple bounces within the island interior at earlier times. As a result, a bright, shell-like structure forms around the magnetic islands, as shown in the final panel. By comparing the two-dimensional temperature map with the final panel of Figure \ref{fig:profile}(b), we confirm that the bright proton-temperature shell lies along the newly reconnected field line separating the high–guide-field island interior from the surrounding low–guide-field region.

However, these observations raise a question: why do electrons not exhibit a bright temperature shell around the large island while the protons do? To answer this question, the heating rate is calculated for both species.

\begin{equation}
\label{eq:electron_energy_equation}
\frac{dU}{dt}
= E_{\parallel} J_{\parallel}
+ \frac{p_{\perp}}{B}
\left(
\frac{\partial B}{\partial t}
+ \mathbf{u}_E \cdot \nabla B
\right)
+ \left(
p_{\parallel}
+ m n u_{\parallel}^2
\right)
\mathbf{u}_E \cdot \boldsymbol{\kappa}
\end{equation}
where $U$ is the total kinetic energy, $\mathbf{u}_E$ is the $\mathbf{E}\times\mathbf{B}$ drift velocity, and $p_{\parallel}$ and $p_{\perp}$ are the parallel and perpendicular pressures, respectively \citep{dahlin14a}. Note that this a guiding-center equation for particle heating, which describes the particle dynamics in the {\it kglobal} model. The first term on the right-hand side of Equation~(\ref{eq:electron_energy_equation}) represents acceleration by the parallel electric field, $E_{\parallel} J_{\parallel}$. The second term corresponds to perpendicular heating associated with conservation of the magnetic moment $\mu$; the quantity in parentheses is effectively $dB/dt$. The third term drives parallel acceleration and arises from the first-order Fermi mechanism through the magnetic curvature term $\mathbf{u}_E \cdot \boldsymbol{\kappa}$, where $\boldsymbol\kappa$ is the curvature of the local magnetic field. The betatron acceleration term is generally smaller than the parallel electric field and curvature-drift terms and is typically a sink during reconnection. We therefore do not discuss the betatron contribution. The other two contributions are shown in Figure~\ref{fig:HeatingPower} for both electrons and protons. Both species exhibit similar behavior in the curvature-drift term, which transitions from weak to strong heating as the low–guide-field region reaches the X-point. 

In contrast, the parallel electric field heating differs markedly between electrons and protons. For electrons, the parallel electric field contribution opposes the Fermi-reflection heating, whereas for protons it acts constructively with it. This difference arises primarily due to the strong electric field in the exhaust region, which is known to trap electrons and inhibit their escape \citep{egedal09a,haggerty15a}. In the {\it kglobal} model, a self-generated electric field holds back the expansion of hot electrons along a magnetic field. This behavior was previously benchmarked in simulations that compared the field line expansion of a region of high temperature with {\it kglobal} and a PIC model \citep{arnold19a}. At late time, Fermi reflection drives strong electron and proton heating downstream from the dominant X-point ($x/L_0=-2.4$ in Figure \ref{fig:te_ti_2cols_tight} at $t/\tau_A=17$). Electrons expanding away from this hot region lose energy by working against the opposing $E_\parallel$. In contrast, protons undergoing a similar expansion gain energy from that same $E_\parallel$. As a result, the electron temperature map in Figure~\ref{fig:te_ti_2cols_tight}(a) does not exhibit an extremely hot shell, whereas a prominent shell structure is present for protons. A more detailed investigation of electron escape and associated energy loss in the context of solar flares will be presented in a future paper.

Figure~\ref{fig:emission_stack}(a) shows the time evolution of the average electron and proton temperatures within the outermost magnetic separatrix. Between $t/\tau_A = 5$ and $t/\tau_A = 14.5$, both temperatures increase gradually as reconnection develops. 
After $t/\tau_A = 14.5$, the weak–guide-field region reaches the reconnecting layer, which leads to a rapid increase in the strength of Fermi reflection. As a result, the proton temperature increases rapidly after this time. In contrast, the rate of electron temperature increase is visible but more modest.  The difference in the thermal heating sensitivity to the guide field of electrons and protons was also found in previous studies \citep{arnold21a,yin24b}.  The dashed line shows the average temperature of the electrons and protons, which assumes that both have reached thermal equilibration through Coulomb collisions. The equipartition time for a system in which the electron temperature and density approximately equal the proton temperature and density is:

\begin{equation}
\tau_{e|i}\approx 52  \,  \left(\frac{10^{10}\text{ cm}^{-3}}{n_i}\right)  \, \left(\frac{20}{\lambda}\right)  \, \left(\frac{T}{10^3\text{ eV}}\right)^{3/2} \quad \mathrm{sec}, 
\label{eq:eqtime}
\end{equation}
where $n$ is the electron density, $T$ is the temperature, and $\lambda$ is the Coulomb logarithm. For typical coronal values we obtain $\tau_{e i}\approx 52$ s, which is much shorter than the duration of most reported hot-onset events. Therefore, electron–proton energy exchange can be effective during the hot-onset phase, and the dashed curve provides a useful proxy for comparison with observationally inferred temperatures.  Our simulation yields a temperature of approximately $0.06 \, m_i C_A^2$ during the hot-onset phase, which corresponds to 1.08 keV or 12.5 MK based on the normalization discussed in the previous section. Overall, after reconnection initiates, the simulated temperature agrees well with the observed hot-onset range (10–15 MK), and its relatively stationary behavior prior to the high–guide-field regime is consistent with the observations. 


Figure~\ref{fig:emission_stack}(b) shows the fraction of particles in the entire domain with energies exceeding 3 times the peak late-time thermal energy of the heated plasma. This is a rough measure of the fraction of non-thermal particles. The fraction of energetic particles, for both electrons and protons, remains relatively low following the onset of reconnection at $t/\tau_A \approx 4.5$, and then increases rapidly once the low–guide-field region becomes involved at $t/\tau_A \approx  14.5$. Note that the fraction of energetic electrons increases more rapidly than the electron temperature during this time. Thus, while low guide field reconnection only modestly increases the electron temperature, it dramatically increases the production of energetic electrons.  The sharp rise in the energetic particle population marks the onset of the impulsive phase of a solar flare, while the earlier interval corresponds to the hot-onset phase. 

The emission measure is shown in Figure~\ref{fig:emission_stack}(c), where $n_0$ is the characteristic particle density. The emission measure is computed by integrating the squared particle density over the region interior to the outermost separatrix. Because magnetic reconnection heats the plasma trapped within the magnetic islands, the plasma inside the separatrices is identified as a source of thermal soft X-ray emission that is used to calculate the emission measure. It is normalized by $2\pi^3 L_0^3 n_0^2$, corresponding to the total emission measure of the simulation domain at the initial uniform density. The steady increase of the emission measure during the hot-onset phase closely resembles the hot-onset observational signatures \citep{hudson21,Battaglia23}. The increase of volume of the heated plasma causes the increase of the emission measure.

\begin{figure*}[ht!]
\plotone{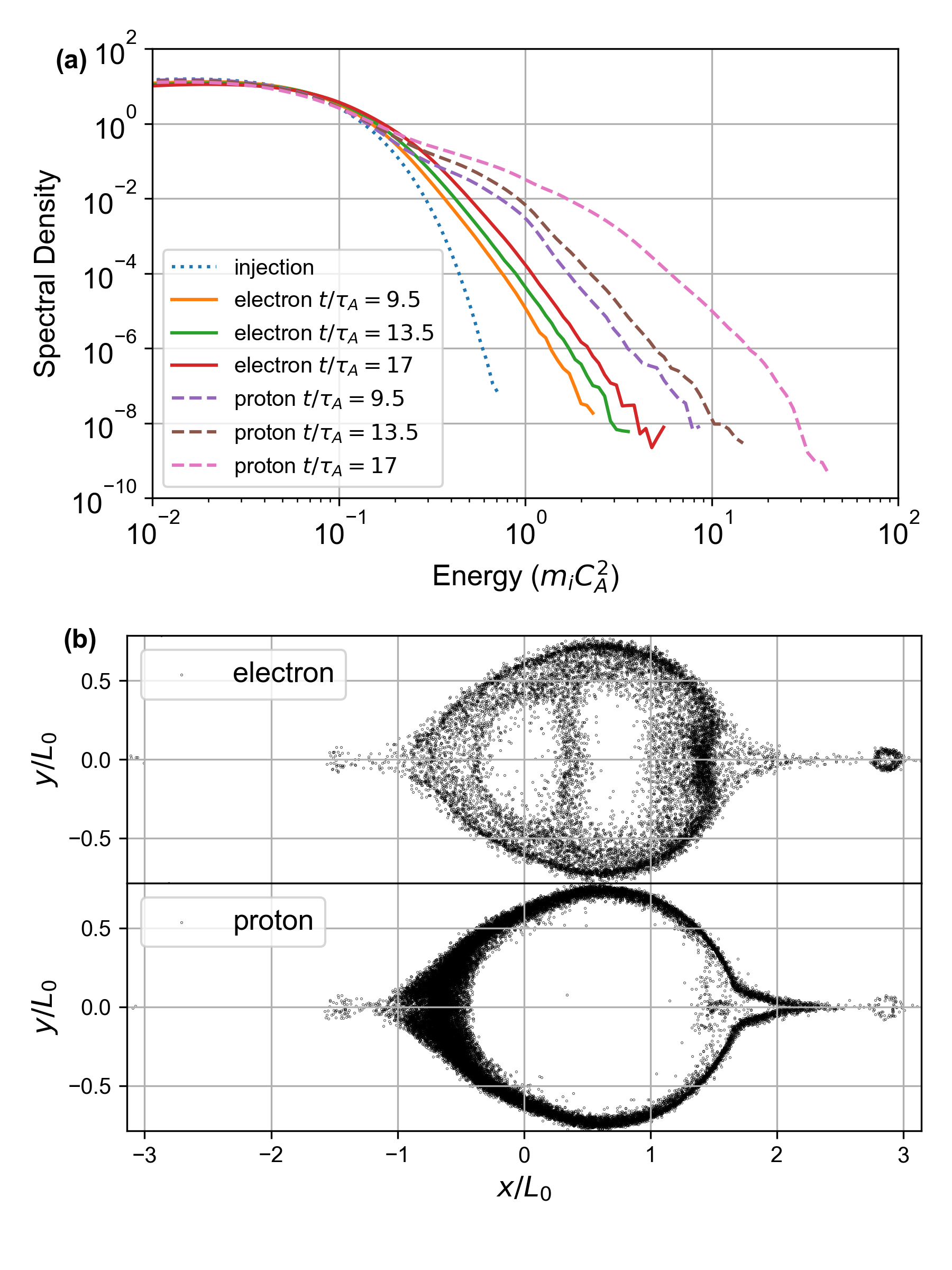}
\caption{In (a):The electron and proton distribution function versus energy. The initial spectrum (blue) for the electron and proton are the same as they share the same temperature. Solid curves are the electron spectrum and dashed curves are proton spectrum at the $t/\tau_A =9.5,13.5$ and $17$. In (b): The locations of electrons with energy exceeding 1.0 $m_iC_A^2$ and protons with energy exceeding 10.0 $m_iC_A^2$.
\label{fig:combined_vertical}}
\end{figure*}
We show the late-time electron and proton energy spectra integrated over the entire simulation domain at several times in Figure~\ref{fig:combined_vertical}(a). At $t/\tau_A = 0$, electrons and protons share the same initial temperature, and their distributions follow the injection spectrum. As the system evolves through the hot-onset phase  ($t/\tau_A=9.5$ and $t/\tau_A = 13.5$), both species exhibit clear energization relative to the initial state. Because these spectra are domain-integrated, they characterize the global outcome of energization and do not uniquely identify the local acceleration sites.

After the low–guide-field region reaches the acceleration layer ($t/\tau_A \approx 14.5$), the spectra develop an increasingly prominent high-energy component. For electrons, the high-energy tail becomes more pronounced and the cutoff energy increases with time, finally reaching a maximum energy $\sim 5m_iC_A^2\sim 90\text{ keV}$. For protons, a harder energetic component emerges at late times ($t/\tau_A=17$), indicating a growing non-thermal proton population. Notably, protons gain substantially more energy in the low–guide-field regime than electrons, which is consistent with earlier simulations with a constant guide field \citep{yin24b,yin25a}. The steep drop at the highest energies is likely attributable to finite-time/finite-size limitations (i.e., the maximum attainable energy within the simulated duration and domain), rather than a physical cutoff established by steady-state losses \citep{yin26a}.

Figure~\ref{fig:combined_vertical}(b) shows the spatial distributions of electrons with energies exceeding $1.0 \:m_i C_A^2$ and protons exceeding $10.0 \:m_i C_A^2$ at $t/\tau_A = 17$. The thresholds are set to highlight the non-thermal components of each species (electrons and protons reach different characteristic energies). These distributions support our interpretation that the newly developed weak guide field at the dominant x-line at late time drives strong local energization, leading to the late time emergence of non-thermal particles. 

\section{Discussion}\label{sec:cite}

We present simulations that suggest that the time evolution of the magnetic guide field, and specifically the transition from a strong to a weak guide field, reproduces key features of hot-onset solar flares. Our {\it kglobal} simulations reveal that during the early stages of reconnection, a strong guide field effectively suppresses Fermi reflection, inhibiting the production of non-thermal particles while sustaining magnetic reconnection that drives thermal heating. This reproduces the key observational signatures of the hot onset: a gradual rise in emission measure and elevated plasma temperatures, about 12.5 MK for typical flare parameters, without the production of non-thermal particles that drive hard X-rays. As the guide field weakens, the system transitions into an efficient particle accelerator, marking the onset of the impulsive phase. This model naturally explains the delayed appearance of non-thermal emissions without requiring separate trigger mechanisms for thermal and non-thermal energy release.

The conventional picture of flare energy release by reconnection fails to account for this distinct delay. In typical simulations with a uniform, weak guide field, Fermi acceleration operates efficiently from the onset, leading to the simultaneous production of thermal and non-thermal populations. This contradicts observations where non-thermal signatures are absent during the early phase of the flare. Our results suggest that the missing ingredient in the standard model is the dynamic evolution of magnetic shear. Such shear evolution has been documented in two-ribbon flares. The hot onset phase of flares emerges naturally as a consequence of the changing guide field strength rather than as a separate precursor process.

As an extension of the present simulations, we plan to add collisional electron-proton temperature equilibrium into {\it kglobal} model since the hot-onset phase lasts longer than the typical equilibration time. The scattering will transfer part of the proton energy to the electron energy. Since the dominant energization mechanism in this study is Fermi reflection (as shown in  Figure \ref{fig:HeatingPower}) and the rate of energy gain from Fermi reflection is proportional to the energy of a particle, we anticipate stronger electron thermal and non-thermal acceleration. Especially in the weak-guide field, protons would transfer a significant amount of energy to electrons, so the stronger Fermi reflection on electrons might produce a harder power-law tail.


The temperature increment of electrons and protons during the hot onset phase in Figure \ref{fig:emission_stack}(a) is only a factor of around two above the initial temperature. This increment is somewhat below the increment seen in observations, However, this is probably a consequence of the initial temperature in our simulation being greater than the preflare coronal temperature. In future simulations we plan to explore how the heating increment during the strong guide field phase depends on initial plasma temperature. 

While the present simulations reproduce the basic characteristics of the hot onset solar flares, the model remains incomplete in that the present simulations are limited to 2D. It is well known that the chaotic magnetic fields in 3D systems can boost the rate of electron energy gain by enabling electrons to escape from flux ropes to sample other energy release sites \citep{dahlin15a,zhang21a}. In 3D the twisted magnetic fields within flux ropes are susceptible to instabilities, such as tearing and kink instabilities, which drive chaotic magnetic fields \citep{Kowal17}. 3D simulations enable the flux-rope kink instability to develop in weak guide field reconnection \citep{zhang21a}, and the kink instability drives field-line chaos which increases the efficiency of Fermi reflection. From an energy perspective, transitioning from 2D to 3D does not significantly alter the total magnetic energy release, but it increases the efficiency of non-thermal energization by enabling particles to experience multiple island contractions \citep{dahlin17a,zhang21a}. Note, however, that the dynamics in 3D does not change the impact of the guide field on particle energization -- strong guide field reconnection in 3D continues to be an inefficient driver of non-thermal particles \citep{dahlin17a}. Consequently, three-dimensional reconnection simulations are expected to yield stronger production of non-thermal particles once reconnection transitions to the weak guide field regime. 

\begin{acknowledgments}
We acknowledge extensive discussions with Dr.\ Hugh Hudson.
Support was provided from NASA grant No.\  
80NSSC20K1813,
and NSF grant No.\  PHY2109083. The simulations were carried
out at the National Energy Research Scientific Computing Center
(NERSC). The data used to perform the analysis and construct
the figures for this paper are preserved at the NERSC High
Performance Storage System and are available upon request.
\end{acknowledgments}

\begin{contribution}

H. Da performed the simulations and the analysis, and wrote the manuscript. All authors contributed to the interpretation of the results and commented on the manuscript.

\end{contribution}

\bibliography{paper}{}
\bibliographystyle{aasjournalv7}

\end{document}